\documentclass[useAMS,usenatbib]{mn2e}

\usepackage{lscape,graphicx,amssymb,aas_macros}
\usepackage{amsmath} 
\usepackage{amssymb} 
\usepackage{graphicx} 
\usepackage{indentfirst} 
\usepackage{pdflscape} 
\usepackage{placeins} 
\usepackage{nicefrac} 
\usepackage{afterpage} 
\usepackage{booktabs} 
\usepackage{color} 
\usepackage{float}

\newcommand{\cmjj}{\mbox{${\rm cm^{-2}}$}}

\newcommand{\etal}{et al.}

\newcommand{\kms}{\mbox{km\ s${^{-1}}$}}

\title[IGM gas accretion onto $z\sim2.8$ \textup{Ly}$\alpha$ emitters]{Probing IGM accretion onto faint Ly$\alpha$ emitters at $z\sim2.8$} 
\author[Zahedy et al.]{Fakhri S. Zahedy$^{1,2}$\thanks{E-mail:
fsz@uchicago.edu}, Michael Rauch$^{2}$, Hsiao-Wen Chen$^{1,3}$, Robert F. Carswell$^{4}$, \newauthor Brian Stalder$^{5}$, Antony A. Stark$^{6}$ \\ \\
$^{1}$Department of Astronomy \& Astrophysics, The University of Chicago, Chicago, IL 60637, USA \\
$^{2}$The Observatories of the Carnegie Institution for Science, 813 Santa Barbara Street, Pasadena, CA 91101, USA \\
$^{3}$Kavli Institute for Cosmological Physics, The University of Chicago, Chicago, IL 60637, USA  \\
$^{4}$Institute of Astronomy, University of Cambridge, Cambridge CB3 0HA, UK \\
$^{5}$LSST, 950 North Cherry Avenue, Tucson, AZ 85719, USA \\
$^{6}$Center for Astrophysics \text{\textbar} Harvard \& Smithsonian, Cambridge, MA 02138, USA}

\begin{document}

\pagerange{\pageref{firstpage}--\pageref{lastpage}} \pubyear{2019}

\maketitle

\label{firstpage}

\begin{abstract}

Observing the signature of accretion from the intergalactic medium (IGM) onto galaxies at $z\sim 3$ requires the detection of faint ($L \ll L^*$) galaxies embedded in a filamentary matrix of low-density ($\rho < 100\ \overline{\rho}$), metal-poor gas ($Z \sim 10^{-2.5} Z_{\odot}$) coherent over hundreds of kpc. We study the gaseous environment of three Ly$\alpha$ emitters (LAEs) at $z=2.7-2.8$, found to be aligned in projection with a background QSO over $\sim 250$ kpc along the slit of a long-slit spectrum.  The lack of detection of the LAEs in deep continuum images and the low inferred Ly$\alpha$ luminosities show the LAEs to be intrinsically faint, low-mass galaxies ($L\lesssim0.1\,L^*$, $M_\mathrm{star}\lesssim 0.1\,M^*$).  An echelle spectrum of the QSO reveals strong Ly$\alpha$ absorption within $\pm200$ \kms\ from the LAEs. Our absorption line analysis leads to \ion{H}{I} column densities in the range of log\,$N\mathrm{(\ion{H}{I})}/\cmjj=16-18$. Associated absorption from ionic metal species \ion{C}{IV} and \ion{Si}{IV} constrains the gas metallicities to $\sim 0.01$ solar if the gas is optically thin, and possibly as low as $\sim0.001$ solar if the gas is optically thick, assuming photoionization equilibrium. While the inferred metallicities are at least a factor of ten lower than expected metallicities in the interstellar medium (ISM) of these LAEs, they are consistent with the observed chemical enrichment level in the IGM at the same epoch. Total metal abundances and kinematic arguments suggest that these faint galaxies have not been able to affect the properties of their surrounding gas. The projected spatial alignment of the LAEs, together with the kinematic quiescence and correspondence between the LAEs and absorbing gas in velocity space suggests that these observations probe a possible filamentary structure. Taken together with  the blue-dominant Ly$\alpha$ emission line profile of one of the objects, the evidence suggests that the absorbing gas is part of an accretion stream of low-metallicity gas in the IGM.

\end{abstract}


\section{Introduction}

The presence of heavy elements in the intergalactic medium (IGM) can be used to probe the relationship between galaxies and the IGM. Absorption lines caused by common metals in the spectra of background QSOs, when observed in sightlines close to foreground galaxies, and, occasionally,  supplemented by observations of nebular emission from the gas, can shed light on how galaxies accrete gas and smaller building blocks from the ambient IGM to fuel star formation, and subsequently eject metals and radiation to their environment. 

Adding metals ab-initio to a cosmological hydro-simulation of the IGM, Haehnelt \etal\ (1996) showed that  typical QSO metal absorbers  can arise during the inflow of pre-enriched gas ($Z\lesssim-2.5\,Z_\odot)$  into $z\sim 3$ galactic potential wells of progenitors ($M_{\rm baryon} \sim 10^{9-10} M_{\odot}$) of galaxies like the Milky Way. The close agreement between the properties of the simulated and observed $z\sim3$ metal absorption lines, even in the absence of any local metal enrichment or galactic feedback (Rauch \etal\ 1997), suggest that metal absorption systems (e.g., \ion{C}{IV}) may provide our earliest observational glimpses of the {\it accretion of gas} onto high redshift galaxies.

The early-enrichment or infall scenario for QSO metal absorbers can be contrasted with a more local origin for the heavy metals, where feedback from recent star formation activity enriches the gaseous environment of a galaxy with metals (the circumgalactic medium, CGM; e.g. Adelberger \etal\ 2003; Steidel \etal\ 2010). However, the existence of long-range outflows able to explain the widespread metal enrichment through local feedback has been questioned (e.g., Kawata \& Rauch 2007; Gauthier \& Chen 2012). Furthermore, the observed low levels of turbulence and spatial small-scale structure in the IGM seen in multiple lines-of-sight to lensed QSOs broadly support the idea that most metal absorbers are seen in stages of inflow as part of the general IGM, unaffected by recent feedback (e.g., Rauch \etal\ 2001a,b). A study of the CGM of Lyman break galaxies has confirmed that the redshift distortions in the velocity field of \ion{H}{I}, \ion{Si}{IV} and \ion{C}{IV} absorption at $z=2-3$ are in fact due to infall, not outflows (Turner \etal\ 2017).

The formation of QSO metal absorbers from infalling IGM gas requires an earlier epoch of widespread metal enrichment in the IGM, the existence of which has been established by detections of ionic metal species at high redshifts currently up to $z\sim 6.8$ (e.g., Songaila 2001; Schaye et al 2003; Simcoe et al. 2004, 2011; Ryan-Weber et al. 2009; Becker et al. 2011; Chen et al. 2017; Cooper \etal\ 2019). Theoretical and empirical arguments also support the idea that the IGM was likely polluted by more widely distributed, low-mass galaxies at earlier times. Gas escapes the shallow potential well of low-mass objects more easily and high-redshift dwarf galaxies occur at high number densities, favoring a widespread pollution of the IGM at early times, aided by the Hubble expansion (e.g., Scannapieco 2005; Porciani \& Madau 2005; Wiersma \etal\ 2010). 

Recent advances in cosmological hydrodynamic simulations have refined predictions for galactic accretion and emphasized the importance of small scale structure embodied by cold accretion streams (e.g., Kere{\v s} \etal\ 2005; Goerdt \etal\ 2010; Faucher-Giguere \& Kere{\v s} 2011; Fumagalli \etal\ 2011; Rosdahl \& Blaizot 2012; Van de Voort \etal\ 2012; Shen \etal\ 2013). Comparisons between observations and the expected accretion features in these cosmological simulations need to rely on statistical arguments, focusing on a subset of properties like column density or metallicity distributions (e.g., Cooper et al, 2015; Hafen et al. 2017). 

In contrast, direct detections of accretion onto individual high-redshift galaxies have remained rare, as they must rely on multiple strands of evidence, essentially requiring two- or three-dimensional information that can only be provided by simultaneous observations of gas and stars in emission. Evidence for gaseous accretion onto galaxies may include kinematic signatures of gas streams attached to and/or moving towards a galaxy, gaseous filaments converging on starburst regions fueled by the inflow, and tidal tails from the simultaneous infall of small halos. One candidate system for filamentary infall onto a high-redshift galaxy was reported by Rauch \etal\ (2011, 2016). That particular system owes its discovery to fluorescent radiation from gaseous features which have exposed to ionizing photons escaping from the starbursting region. However, such a situation may be atypical of galaxies undergoing the cold accretion process, most of which will have too low star-formation rates and presumably masses required to illuminate their gaseous environment.  

\begin{figure} 
\hspace{-0.1in}
\includegraphics[width=87mm]{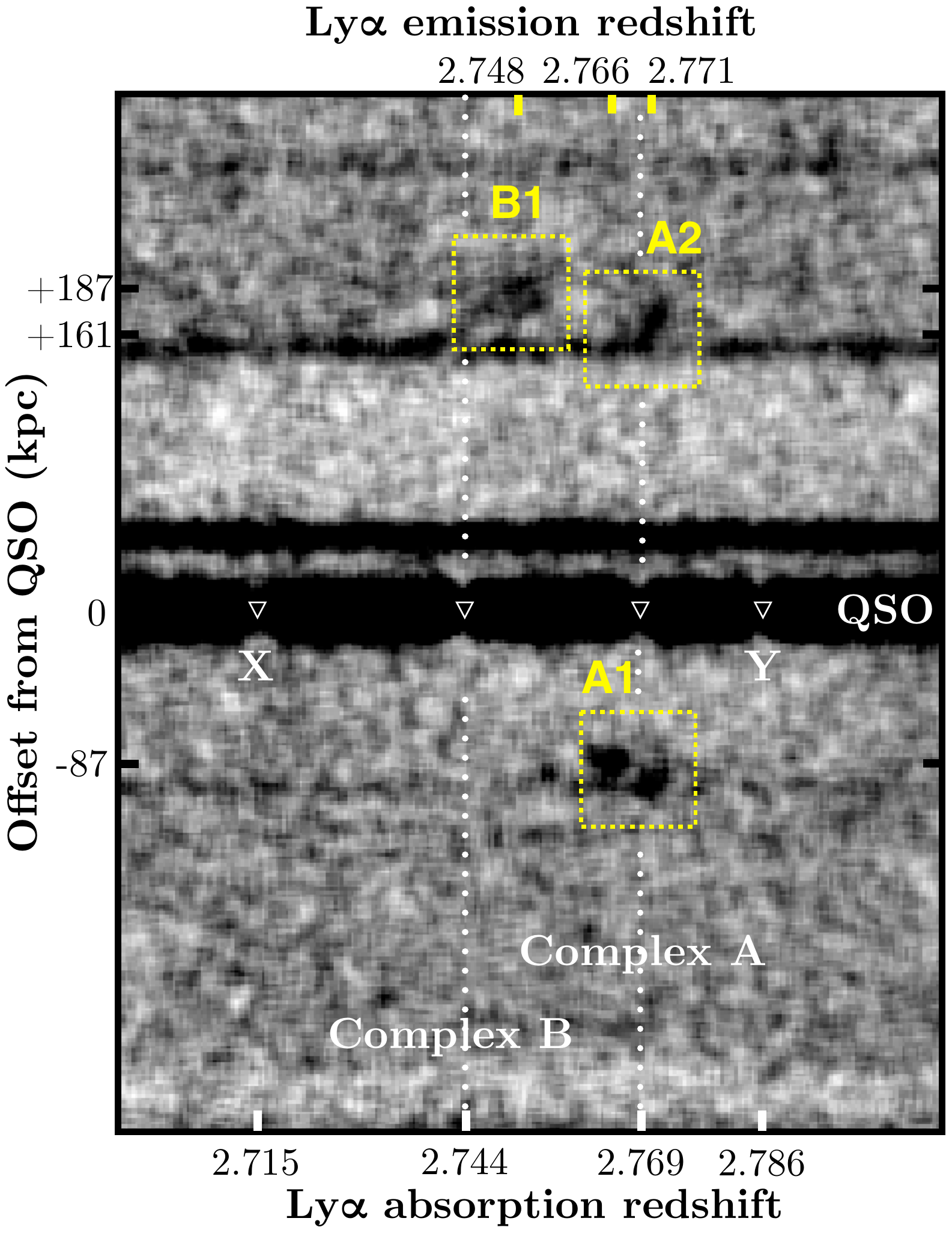}
\vspace{-1.5em}
\caption{Two-dimensional long-slit spectrum centered at QSO DMS 2139$-$0405.  The dispersion axis is in the horizontal direction, covering roughly $4480-4640$ \AA\ in observed wavelength, which corresponds to the location of Ly$\alpha$ transition at $z=2.7-2.8$. The spatial axis is in the vertical direction and covers 75 arcsec from top to bottom, approximately 600 kpc in projected distance. The strong continuum at the center is the QSO, whereas featureless and fainter continua above it are likely lower-redshift objects. Three foreground Ly$\alpha$ emitters (LAEs) are detected at $d<300$ kpc from the QSO (dotted squares; see Rauch \etal\ 2008), with their Ly$\alpha$ emission redshifts indicated on the upper horizontal axis. In addition, the locations of strong Ly$\alpha$ absorption in the QSO continuum are shown with inverted triangles, with their absorption redshifts indicated on the lower horizontal axis. LAEs A1 and A2 are situated at $d=87$ and 161 kpc on opposite sides of the QSO, where a strong Ly$\alpha$ absorption is coincident with both LAEs in redshift space (Complex A, see Figure 3). Another Ly$\alpha$ absorption system (Complex B, see Figure 4) is detected at $d=187$ kpc and a velocity separation $|\Delta v| < 500$ \kms\ from LAE B1. Two additional Ly$\alpha$ absorption complexes (X and Y) are also found at $z=2.7-2.8$ (see discussion of these systems in Appendix A).}

\end{figure}

Observing more typical forms of galactic accretion requires the ability to simultaneously detect fainter galaxies and fainter nebular emission features.
In this paper we will study such a putative accretion environment, by paying a closer look at a grouping of $z\sim2.8$ Ly$\alpha$ emitters (LAEs) previously discovered through a blind spectroscopic long-slit search (Rauch \etal\ 2008, hereafter R08). As shown in Figure 1, this grouping is unique because the LAEs are coincident in redshift with a number of Ly$\alpha$ absorption complexes seen in the sightline of a nearby background QSO, DMS 2139$-$0405. The underlying galaxies of the LAEs are faint, with low inferred star formation rates of $\mathrm{SFR}< 1 \,\mathrm{M_\odot}\,\mathrm{yr^{-1}}$. In particular, one of the gas complexes, Complex A, occurs in a structure spanned by a pair of Ly$\alpha$-emitting galaxies which are situated at projected distances $d=87$ and 161 kpc on opposite sides of the QSO sightline. One of the LAEs shows a relatively rare Ly$\alpha$ emission profile with a dominant blue peak (R08), which is commonly associated with infalling gas (e.g., Verhamme \etal\ 2006). 

The positional information about the LAEs derived from the two-dimensional low resolution spectra of the field, 
and the high-resolution absorption spectra of neutral hydrogen and metal species along the QSO sightline, allow us to study the physical association between the galaxies and the gas, and to test the predictions of the infall scenario for metal absorption systems against the observed gas column density, chemical abundance, ionization state, and kinematics. The low luminosities and likely low masses of the LAEs also enable us to examine the validity of the late-time outflow scenario and the CGM paradigm in a more typical environment of the cosmic web far from massive starburst galaxies. 

This paper is organized as follows. Section 2 presents the observational dataset  and corresponding data reduction. We describe the properties of the foreground LAEs in Section 3. In Section 4, we present the absorption-line measurements, describe the absorption line complexes near the LAEs, and investigate the ionization and chemical properties of the gas. Finally, we discuss the interpretations of our findings in Section 5 and summarize our study in Section 6.  Throughout this paper, we adopt a $\Lambda$ cosmology of $\Omega_{\rm M}=0.3$ and $\Omega_\Lambda = 0.7$, with a Hubble constant of $H_0 = 70 \ {\rm km} \ {\rm s}^{-1}\ {\rm Mpc}^{-1}$. All magnitudes are reported in the AB system.

\section[]{Observations}

To characterize the properties of the absorbing gas and investigate its connection to the unique grouping of $z\sim2.8$ LAEs, we carried out high-resolution echelle spectroscopy of the background QSO DMS 2139$-$0405 ($z_\mathrm{QSO}=3.32$). In addition, we obtained new optical and near-infrared broadband imaging data of the field around the QSO in order to search for continuum counterparts to the LAEs. Here we describe the observations and data reduction procedures.

\subsection[]{QSO echelle spectroscopy}
We obtained high-resolution echelle spectra of QSO DMS 2139$-$0405 (right ascension $21^\mathrm{h}41^\mathrm{m}39.1^\mathrm{s}$, declination $-03^{\circ}51'42.57''$ in the J2000 epoch) with the MIKE echelle spectrograph (Bernstein \etal\ 2003) on the Magellan Clay Telescope over two nights in 2016 September. A 1-arcsec slit and 3 $\times$ 3 binning during readout was chosen for the observations, delivering a spectral resolution of FWHM $\approx 10-12$ \kms\ over a continuous wavelength coverage from 3350 \AA\ to 9300 \AA. The mean seeing over the observing period was $0.5-0.7$ arcsec. The total integration time of  the MIKE observations was 26400 s, comprising eight individual exposures of equal duration. We reduced the MIKE spectra using an \textsc{idl}-based custom data reduction software. For each individual exposure, the software performed optimal extraction of the QSO spectrum in each echelle order, using a Gaussian weighting scheme that matched the observed QSO spatial profile. Observations of a spectrophotometric standard star taken during the same night as the observations were utilized to determine the response function, which was then used to perform relative flux calibrations of the QSO spectrum. The individual echelle orders were combined for each exposure, and the different exposures were added to form a single spectrum, which was then continuum normalized using a low-order polynomial function that excluded regions of strong absorption. The final reduced spectrum of DMS 2139$-$0405 is characterized by a signal-to-noise ratio of S/N $\approx 10-15$ per resolution element at $\lambda>4000\,$\AA. 

\begin{table}
\begin{center}
\caption{Journal of imaging observations of the field around QSO DMS\,2139$-$0405}
\vspace{-0.5em}
\label{tab:Imaging}
\resizebox{3.4in}{!}{
\begin{tabular}{ccccrc}\hline
Telescope & Instrument 	& Filter & \multicolumn{1}{c}{Exp.}		& \multicolumn{1}{c}{Date} & $5\sigma$ Limiting  \\	
 		  		&&       & \multicolumn{1}{c}{Time (s)}     &  & Mag. (AB)	       \\\hline \hline
Magellan Clay		&PISCO 	&$g$ 	& 	3300		&  2016 Nov	&27.6 \\
				&PISCO 	&$r$ 	& 	3300		&  2016 Nov	&27.2   \\
  				&PISCO 	&$i$ 		& 	3300		&  2016 Nov	&26.7   \\  
				&PISCO 	&$z$ 	& 	3300		&  2016 Nov	&25.9   \\ \hline
Magellan Baade	&FourStar 	&$K_s$ 	& 14120		&  2016 Aug	&25.1  \\  \hline 
Keck I			&LRIS 	&$V$ 	& 	1680		&  1998 Jun	&27.1  \\
				&LRIS 	&$I$ 		& 	4050		&  1998 Aug	&26.2   \\  

\hline
\end{tabular}}
\end{center}
\end{table}

\subsection[]{Imaging observations}

Optical imaging data of the field around DMS 2139$-$0405 were obtained using the
Parallel Imager for Southern Cosmology Observations (PISCO; Stalder \etal\ 2014) on the Magellan Clay Telescope in 
2016 October. PISCO is a multi-band imager which provides simultaneous broadband coverage in
$g$-, $r$-,$i$-, and $z$-bands. The observations consisted of a series of exposures 300 s in length, with a total integration
time of 3300 s taken under a mean seeing of $ 0.9-1.0$ arcsec. The raw PISCO data were reduced using standard \textsc{iraf}-based routines.  
The photometric zero points of the PISCO imaging data were determined using field stars observed in the Sloan Digital Sky
Survey (SDSS; York et al. 2000). We supplement our PISCO imaging data with archival $V$- and $I$-band images of the field taken with LRIS (Oke \etal\ 1995) on the Keck-I Telescope, which were previously described in R08. The LRIS $V$- and $I$-band images are characterized by a mean seeing of $ 0.7-0.9$ arcsec. 

In addition to the optical images, we obtained near-infrared (NIR) $K_s$-band observations of the field with FourStar (Persson \etal\ 2008) on the Magellan Baade
Telescope in 2016 August. The field was observed for a total of 14120 s, under excellent seeing conditions of $0.5-0.8$ arcsec. To prevent saturation
due to high sky and thermal backgrounds in the image, the observations consisted of a sequence of short 20 s exposures with a
10-arcsec random dither applied between successive exposures. The FourStar data were reduced 
using an \textsc{iraf}-based routine which performed sky subtraction and flat-fielding using super sky frames created on a rolling basis. We used field stars
observed with the Two Micron All Sky Survey (2MASS; Skrutskie 2006) to determine the photometric zero point of the $K_s$-band image. 

We summarize the available imaging data of the field around QSO DMS 2139$-$0405 
in Table 1, where we also show the mean 5$\sigma$ limiting magnitudes measured over a 1 arcsec sky aperture for each bandpass. The limiting magnitudes range from 25.1 mag for NIR $K_s$ band to deeper than 27 mag for the optical $g$-, $V$-, and $r$-bands. In Figure 2, we present a false-color composite image of the field around QSO DMS\,2139$-0405$. The composite image was created using the $V$-, $I$-, and $K_s$-band images for blue, green, and red, respectively.

\begin{table*}
\begin{center}
\caption{Properties of LAEs at $d<300$ kpc from QSO DMS\,2139$-$0405 ($z_\mathrm{QSO}=3.32$)}
\vspace{-0.5em}
\label{tab:Imaging}
\resizebox{6.9in}{!}{
\begin{tabular}{cccrccccccccccccc}\hline
ID	&	Alt. ID$^a$ 	& 	${\Delta\mathrm{RA}}^b$ 		& 	${\Delta\mathrm{Dec}}^b$ & $z_\mathrm{peak}^c$&  $z_\mathrm{sys}^d$ 	&$d^e$&  \multicolumn{1}{c}{ ${F_\mathrm{Ly\alpha}}$ }	& \multicolumn{7}{c}{Photometry$^ f$}&$\mathrm{SFR_{UV}}^g$& $\mathrm{SFR_{Ly\alpha}}^h$ 	\\\cline{9-15}	
	& 		&	(arcsec)		&    	(arcsec)		   	&  		& & (kpc)& \multicolumn{1}{c}{($10^{-18}\, \mathrm{erg\,s^{-1}\,cm^{-2}}$)}					&AB($g$)	& AB($V$)&AB($r$)	&AB($i$)	& AB($I$) &AB($z$) &AB($K_s$)&$(\mathrm{M_\odot}\,\mathrm{yr^{-1}})$&$(\mathrm{M_\odot}\,\mathrm{yr^{-1}})$	 		       \\\hline \hline
A1	&	15	&$-1.1$	&$-10.9$	&2.7659   &2.7699     &87	&$2.67\pm0.41$	&$>27.1$	& $>26.3$	&$>26.6$ 	&$>26.0$	& $>25.5$	& $>25.3$	&$>24.5$&$<3.2$&0.1 \\
A2	&	37	&$+2.1$	& $+20.4$	&2.7713   &2.7691    &161	&$3.27\pm0.57$ &$>27.1$	&$>26.3$&$>26.6$&$>26.1$& $>25.5$	& $>25.4$	&$>24.5$&$<3.0$&0.1	 	\\
B1	&	36	&$+2.5$	&$+23.6$	&2.7483   &2.7461    &187	&$3.46\pm0.52$ &$>27.1$	& $>26.3$	&$>26.6$&$>26.1$	& $>25.5$	& $>25.4$	&$>24.5$&$<3.0$&0.1		\\
1	&	12	&$+1.0$	&$+9.8$	&3.3300&3.3275&73	&$3.36\pm0.35$ &$>27.0$	& $>26.3$	&$>26.6$ 	&$>26.1$	& $>25.5$	& $>25.3$	&$>24.5$&$<4.1$&0.2	\\
2	&	33	&$+0.6$	&$+4.7$	&3.2646&3.2621&36	&$3.50\pm0.34$ &$>26.9$	& $>26.3$	&$>26.5$ 	&$>26.0$	& $>25.5$	& $>25.3$	&$>24.4$&$<4.4$&0.2	\\
3	&	16	&$-1.7$	&$-16.9$	&3.3189&3.3164&127	&$3.06\pm0.37$		 	&$>27.2$	& $>26.3$	&$>26.6$ 	&$>26.1$	& $>25.5$	& $>25.3$	&$>24.5$&$<4.1$&0.2 	\\

\hline
 \multicolumn{17}{l}{\bf Notes} \\
 \multicolumn{17}{l}{$^a$ Original ID from R08.}\\
 \multicolumn{17}{l}{$^b$ Offset from the QSO position in J2000, $\mathrm{(RA, Dec.)}$=($21^\mathrm{h}41^\mathrm{m}39.1^\mathrm{s}$, $-03^{\circ}51'42.57''$)}\\
 \multicolumn{17}{l}{$^c$ Peak emission redshift of the LAE, from R08.}\\
 \multicolumn{17}{l}{$^d$ The adopted LAE systemic redshift, calculated by applying an offset of $-174\, (+316$)  \kms\ to $z_\mathrm{peak}$ for redshifted (blueshifted) Ly$\alpha$ line peak, based on Hashimoto \etal\ (2015; see \S 3).}\\
 \multicolumn{17}{l}{$^e$ Projected distance from the QSO. Note that LAEs 1, 2, and 3 are excluded from this study because they are situated at $\Delta v > -10000$ \kms\ from the QSO redshift, $z_\mathrm{QSO}=3.32$ (see \S\ 3).}\\
 \multicolumn{17}{l}{$^f$ Measured in a 3 arcsec aperture centered at the nominal sky coordinates of the object. The upper limits are 3$\sigma$.}\\
 \multicolumn{17}{l}{$^g$ Calculated from the upper limit on the $i-$band flux, using the near-ultraviolet (NUV) star-formation rate (SFR) calibrator from Kennicutt \& Evans (2012)}\\
 \multicolumn{17}{l}{$^h$ Calculated from the Ly$\alpha$ line luminosity of the LAE,  by converting the local H$\alpha$ SFR calibrator from Kennicutt \& Evans (2012) to one for Ly$\alpha$, assuming case B recombination (see \S 3).}\\
  
\end{tabular}}
\end{center}
\end{table*}

\begin{figure} 
\includegraphics[width=85mm]{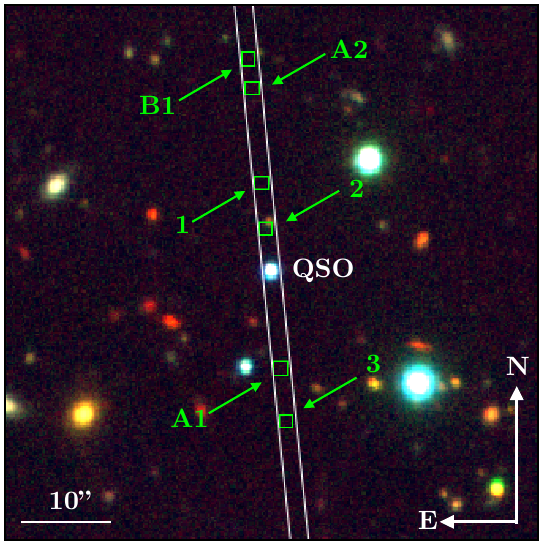}
\caption{False-color composite image of the field around QSO DMS\,2139$-0405$, created using the  $V$-, $I$-, and $K_s$-band images for blue, green, and red, respectively. The QSO is the blue point source at the center of the image. The slit position and orientation of the deep spectroscopic observations reported in R08 are shown. The position of LAEs detected along the slit at $d<300$ kpc from the QSO is shown with green squares and labeled by their IDs in Table 2.}
\label{Figure}
\end{figure}

\section[]{Foreground LAE\MakeLowercase{s} at $\mathbf{\MakeLowercase{\mathit{d}}<300}$ \MakeLowercase{kpc} from DMS\,2139$-$0405}

To study the gaseous environment of high-redshift LAEs, we focus on foreground LAEs within $d<300$ kpc from the sightline of background QSO DMS\,2139$-$0405. Systematic surveys of gas absorption around $z=2-3$ galaxies have shown enhanced gas absorption at $d\lesssim300$ kpc (e.g., Rudie \etal\ 2012). As shown in Figure 2, six LAEs from the R08 study are detected within $d<300$ kpc from the background sightline. To avoid confusion with possible outflows from the background QSO (e.g., Wild \etal\ 2008), we further limit our study to LAEs that are situated at a line-of-sight velocity difference of $\Delta v <-10 000$ \kms\ from the background QSO. Three Ly$\alpha$-emitting objects, LAEs 1, 2, and 3, which are located at $z\sim3.3$ or less than a few\,$\times\,1000\,$\kms\ in velocity separation from the QSO, are excluded due to this requirement. 

There are three foreground LAEs at $d<300$ kpc from the QSO sightline. Two of these objects, LAE A1 at peak redshift $z_\mathrm{peak}= 2.7659$ and LAE A2 at $z_\mathrm{peak}= 2.7713$, are closely situated in redshift space with their Ly$\alpha$ emission peaks separated by merely $\Delta v \approx 400$ \kms. In addition, both LAEs are situated on opposite sides of the QSO at a projected separation of 248 kpc, with the QSO sightline probing the LAE A1 at $d=87$ kpc and  LAE A2 at $d=161$ kpc (top-left panel of Figure 3). A third emitter, LAE B1, occurs at $z_\mathrm{peak}=2.7483$, and is probed by the background QSO at $d=187$ kpc (top-left panel of Figure 4). 

\begin{table*}
\begin{center}
\caption{Properties of $z=2.7-2.8$ absorption complexes along sightline DMS\,2139$-$0405}
\vspace{-0.5em}
\label{tab:Integrated properties}
\resizebox{7.in}{!}{
\begin{tabular}{@{\extracolsep{4pt}}cccccccccccc@{}}\hline

Complex &$z_\mathrm{abs}$ 					& \multicolumn{2}{c}{\ion{H}{I} $\lambda$1215}& \multicolumn{2}{c}{\ion{Si}{IV} $\lambda$1393}& \multicolumn{2}{c}{\ion{Si}{IV} $\lambda$1402}& \multicolumn{2}{c}{\ion{C}{IV} $\lambda$1548}&\multicolumn{2}{c}{\ion{C}{IV} $\lambda$1550}\\ \cline{3-4}\cline{5-6}\cline{7-8}\cline{9-10}\cline{11-12}	
&								& $W_r$			& [$v_\mathrm{min},v_\mathrm{max}$]& $W_r$			& [$v_\mathrm{min},v_\mathrm{max}$] & $W_r$			& [$v_\mathrm{min},v_\mathrm{max}$] & $W_r$			& [$v_\mathrm{min},v_\mathrm{max}$] & $W_r$			& [$v_\mathrm{min},v_\mathrm{max}$]           \\ 
&								& $\mathrm{(\AA)}$	&  (km\,s$^{-1}$)& $\mathrm{(\AA)}$	&  (km\,s$^{-1}$)& $\mathrm{(\AA)}$	&  (km\,s$^{-1}$)& $\mathrm{(\AA)}$	&  (km\,s$^{-1}$)& $\mathrm{(\AA)}$	&  (km\,s$^{-1}$)	\\ \hline \hline

A&  $2.7691$	&  $1.24\pm0.02$	&  $[-225,+225]$	&  $0.05\pm0.01$	& $[-100,+100]$ 	& $0.03\pm0.01$&  $[-100,+100]$ &  	  $0.35\pm0.02$	&  
$[-100,+100]$	&  $0.26\pm0.02$	&  $[-100,+100]$	 	\\ 
B& $2.7440$	&  $1.99\pm0.03$	&  $[-255,+645]$	&  ...	& ... 	& $<0.02$&  $[-75,+65]$ &  	  $0.09\pm0.02$	&  $[-75,+65]$	&  $0.07\pm0.02$	&  $[-75,+65]$	 	\\ 

\hline
\end{tabular}}
\end{center}
\end{table*}

Because the shape of Ly$\alpha$ emission line is highly sensitive to details of the radiative transfer (e.g., Verhamme \etal\ 2006), it is likely that the systemic redshifts of the LAEs, $z_\mathrm{sys}$, are different from their measured Ly$\alpha$ peak redshifts, $z_\mathrm{peak}$. Previous studies found that a large majority of Ly$\alpha$-selected galaxies at $z\sim2-3$ exhibit asymmetric Ly$\alpha$ emission line profiles, with a main peak that is typically redshifted from the systemic redshift by greater than $100$ \kms\ (e.g., Erb \etal\ 2014; Hashimoto \etal\ 2015; Trainor \etal\ 2015). Using a sample of faint LAEs, Hashimoto \etal\ (2015) found a mean offset of $\Delta v_\mathrm{red}=+174$ \kms\ for the main red peak relative to systemic.  To estimate the systemic redshifts of the LAEs, we adopt the mean offset from Hashimoto \etal\ (2015) and apply it to the peak redshifts of LAEs A2 and B1. For LAE A1, which exhibits a dominant {\it blue} peak in its Ly$\alpha$ emission profile (R08), we estimate its systemic redshift by adopting an offset of $\Delta v_\mathrm{blue}=-316$ \kms, which is the mean offset Hashimoto \etal\ (2015) found for blueshifted Ly$\alpha$ emission line peaks. 

The observed Ly$\alpha$ line fluxes of the three LAEs are $F_\mathrm{Ly\alpha}=  (2.7\pm0.4)\times10^{-18}, (3.3\pm0.6)\times10^{-18}$, and $(3.5\pm0.5)\times10^{-18}\, \mathrm{erg\,s^{-1}\,cm^{-2}}$ for LAEs A1, A2, and B1 (R08). These observed line fluxes correspond to Ly$\alpha$ luminosities of $L_\mathrm{Ly\alpha}=(1.7-2.2)\times10^{41}\, \mathrm{erg\,s^{-1}}$.
The inferred Ly$\alpha$ luminosities of the three LAEs are a factor of $20-30$ lower than the characteristic luminosity of the LAE luminosity function at  $z\sim3$ (e.g., Cassata \etal\ 2011; Konno \etal\ 2016; Drake \etal\ 2017), which indicate that these LAEs are intrinsically faint compared to the general LAE population at the same epoch.\footnote{This conclusion would stand even if the observations were subject to significant slit losses of $\sim$50\% (e.g., R08; Cassata \etal\ 2011).} 

The available multi-band imaging data of the field around DMS\,2139$-$0405 allows us to search for continuum light counterparts of these faint LAEs. We infer the positions of the LAEs on the sky based on the reported long-slit position and orientation for the R08 observations as well as the spatial position of the Ly$\alpha$ emissions in the long-slit spectrum from R08. For each bandpass, we then measure the total flux contained within a 3-arcsec diameter aperture centered at the inferred position of each LAE. The choice of aperture size is guided by the width of the long slit used in R08. We find that none of the three LAEs is detected in continuum light at 3$\sigma$ or higher significance level, in all broadband filters. The lack of significant detection in any broadband image persists if a smaller (2 arcsec) or larger (5 arcsec) aperture is adopted instead. 

The lack of detection of continuum light from all three LAEs ($r>26.6$, $K_s>24.5$) indicates that these LAEs are faint and low-mass galaxies. At $z=2.7-2.8$, the upper limit on $r$-band flux corresponds to a rest-frame FUV absolute magnitude of  $M\mathrm{(1700\,\AA)}\gtrsim-18.7$, which is equivalent to $L\lesssim0.1\,L^*$ (e.g., Reddy \etal\ 2008). Using this limit on FUV absolute magnitude, we constrain the total stellar masses of the LAEs to $M_\mathrm{star}\lesssim10^9\,\mathrm{M_\odot}$, based on the inferred mean mass-to-light ratio and stellar mass function as this epoch (e.g., Reddy \etal\ 2012; Marchesini \etal\ 2009). The inferred low masses of the three LAEs are consistent with the low mean dark matter halo mass ($M_\mathrm{h}\sim10^{11}\,\mathrm{M_\odot}$) of $z\sim3$ LAEs inferred from clustering analyses (e.g., Bielby \etal\ 2016).

Next, we estimate the star-formation rate (SFR) of the LAEs. At $z\approx2.8$, the effective wavelength of the $i$-band filter corresponds to a rest-frame wavelength of $\approx 2050$ \AA\ in the near-ultraviolet (NUV). By converting the $i$-band flux limits of the LAEs to their corresponding upper limits on specific luminosity in the NUV,  and then adopting the local NUV star-formation rate relation from Kennicutt \& Evans (2012, equation 12), we estimate upper limits on unobscured SFR of $\mathrm{SFR_\mathrm{UV}\lesssim 3.0}\,\mathrm{M_\odot}\,\mathrm{yr^{-1}}$  for the three LAEs.

An independent estimate of the star-formation rate of the LAEs can also be obtained by assuming that the observed Ly$\alpha$ emission is powered by underlying star formation activity in the galaxies. If we also assume case-B recombination (Brocklehurst 1971), the Kennicutt \& Evans (2012) SFR relation for H$\alpha$ can be converted into one for Ly$\alpha$, following $\mathrm{SFR_\mathrm{Ly\alpha}}\,(\mathrm{M_\odot}\,\mathrm{yr^{-1}})=6.2\times10^{-43}\,L_\mathrm{Ly\alpha} \,( \mathrm{erg\,s^{-1}})$. Applying this relation to the observed Ly$\alpha$ luminosities gives unobscured SFRs of $\mathrm{SFR_\mathrm{Ly\alpha}\approx 0.1}\,\mathrm{M_\odot}\,\mathrm{yr^{-1}}$ for the three LAEs.  

In Table 2, we summarize the photometric and spectroscopic properties of LAEs A1, A2, and B1. For each object, we report its angular offset from the QSO, observed peak Ly$\alpha$ redshift, estimated systemic redshift, total Ly$\alpha$ line flux, projected distance from the QSO sightline, 3$\sigma$ upper limits on the broadband fluxes, and the inferred SFRs. For completeness, we also include the same information in Table 2 for LAEs 1, 2, and 3, which are located at $z\sim3.3$ or less than a few\,$\times\,1000\,$\kms\ in velocity separation from the QSO.

\section[]{ Analysis and Results}

The high-resolution echelle spectrum of DMS 2139$-$0405 enables a detailed study of gas in proximity to the three foreground LAEs. As previously mentioned in \S 1, a number of Ly$\alpha$ absorption complexes at $z=2.7-2.8$ are found in the low-resolution spectrum of the QSO (Figure 1). In particular, Complex A and Complex B are coincident in redshift space with the three foreground LAEs situated at $d<300$ kpc (see Figure 1 and \S 3). In this section, we describe the observed gas properties in the vicinity of these $z=2.7-2.8$ LAEs, based on our analysis of the high-resolution echelle spectrum of DMS 2139$-$0405. 

\subsection[]{Absorption Analysis}

At the LAE redshifts, our MIKE spectrum of DMS 2139$-$0405 provides coverage of the
\ion{H}{I} Ly$\alpha$ $\lambda$1215 line and prominent ionic metal transitions \ion{C}{IV} $\lambda\lambda$1548,1550 and \ion{Si}{IV} $\lambda\lambda$1393,1402 doublets.\footnote{Note that while the MIKE data cover all higher order Lyman transitions, the presence of a higher-redshift Lyman-limit system results in low transmitted flux for rest wavelengths $\lambda_\mathrm{rest}\lesssim1050\,$\AA\ at the LAE redshifts. Consequently, only Ly$\alpha$ $\lambda$1215 has sufficient data quality for analysis.} To characterize the absorbers, we first measure the total rest-frame equivalent widths for transitions \ion{H}{I} Ly$\alpha$ $\lambda$1215, 
\ion{C}{IV} $\lambda\lambda$1548,1550, and \ion{Si}{IV} $\lambda\lambda$1393,1402 absorption lines. When a given ionic transition is not detected, we use the error array to estimate the 2$\sigma$ upper limit on the absorption equivalent width,  integrated over a spectral window that is twice the FWHM of the corresponding \ion{H}{I} line. The equivalent width measurements are summarized in Table 3, where for each transition we present the absorption redshift $z_\mathrm{abs}$, the rest-frame equivalent width $W_r$, and the velocity interval over which the absorption equivalent width is integrated. 

To further investigate the gas properties, we perform a Voigt profile fitting analysis using the \textsc{vpfit} package (Carswell \& Webb 2014), in order to constrain the centroid redshift, $z_c$, gas column density, log\, $N_c$, and Doppler parameter, $b_c$, of individual absorbing components. We report the results of the Voigt profile fitting in Table 4, and present the best-fit model absorption profiles for both absorption complexes along with the data in Figure 3 and Figure 4. Uncertainties in Table 4 represent the estimated 1$\sigma$ (68\%) confidence level from the fitting analysis. When no \ion{C}{IV} or \ion{Si}{IV} absorption is detected, we estimate the 2$\sigma$ upper limit on the absorption equivalent width of the strongest available transition of the species, which is calculated over a velocity window twice the full-width-at-half-maximum (FWHM) of the corresponding Ly$\alpha$ component. Then, we calculate the corresponding 2$\sigma$ upper limit on that component's column density assuming the gas is optically thin. For saturated Ly$\alpha$ components, we also estimate the range of allowed log\,$N_c$ and $b_c$  at the 95 percent confidence level, based on a likelihood contour constructed from a grid of $\chi^2$ values from comparing Voigt profile models and observations. 

\begin{table}
\begin{center}
\caption{Voigt profile fitting results for $z\sim2.8$ absorption complexes associated with LAEs}
\vspace{-0.5em}
\label{tab:Imaging}
\resizebox{3.4in}{!}{
\begin{tabular}{cccc}\hline
Species		& $z_c$ 					& log\,$N_c$/\cmjj		&$b_c$\\	
 		&								&    		   	& (km\,s$^{-1}$)            \\\hline \hline

\multicolumn{4}{c}{Complex A at $z=2.7691$}\\\hline

\ion{Si}{IV}	&  $ 2.76883\pm0.00008$	&  $12.11\pm0.24$	&    $7.9\pm8.6$	\\	
\ion{C}{IV}		&   $2.76893\pm0.00001$	&  $13.70\pm0.04$	&    $12.5\pm1.6$	\\	
\ion{H}{I}		&   $2.76914\pm0.00001$	&  $17.1\pm0.3    $	&     $57 \pm3   $	\\
\ion{Si}{IV}	&   $2.76926\pm0.00004$	&  $12.11\pm0.21$	&     $4.7\pm8.9$	\\
\ion{C}{IV} 	&   $2.76933\pm0.00002$	&  $13.10\pm0.24$	&     $3.4\pm4.6$	\\
\ion{C}{IV} 	&   $2.76959\pm0.00001$	&  $13.81\pm0.13$	&     $6.3\pm1.6$	\\
\ion{Si}{IV}	&   $2.76962\pm0.00002$	&  $12.47\pm0.11$	&     $6.5\pm4.3$	\\
\ion{C}{IV} 	&   $2.77002\pm0.00002$	&  $13.53\pm0.05$	&   $12.8\pm2.2$	\\
\ion{Si}{IV}	&   $2.77014\pm0.00005$	&  $12.15\pm0.20$	&     $6.0\pm7.3$	\\ \hline

\multicolumn{4}{c}{Complex B at $z=2.7440$}\\\hline

H \,I      	&   $2.7418  \pm  0.0002$	&  $12.98\pm0.29$	&    $25.9\pm16.6$		\\ 
\ion{C}{IV}      	&  ...						&  $<12.8$	&    ...					\\ 
\ion{Si}{IV}      	&  ...						&  $<12.5$	&    ...					\\ 
\ion{C}{IV}     	&   $2.74398\pm0.00006$	&  $13.53\pm0.07$	&    $40.4\pm7.2$		\\ 	
\ion{H}{I}      	&   $2.74400\pm0.00002$	&  $17.3^{+0.3}_{-0.5}$&    $52^{+5}_{-2}	$		\\ 
\ion{Si}{IV}      	&  ...						&  $<12.7$	&    ...					\\ 
\ion{H}{I}      	&   $2.74755\pm0.00003$	&  $13.36\pm0.06$	&    $21.8\pm3.0$		\\ 
\ion{C}{IV}      	&  ...						&  $<12.8$	&    ...					\\ 
\ion{Si}{IV}      	&  ...						&  $<12.5$	&    ...					\\ 
\ion{H}{I}      	&   $2.74915\pm0.00002$	&  $14.56\pm0.06$	&    $51.5\pm3.1$		\\ 
\ion{C}{IV}      	&  ...						&  $<12.9$	&    ...					\\ 
\ion{Si}{IV}      	&  ...						&  $<12.7$	&    ...					\\ 
\ion{H}{I}      	&   $2.75127\pm0.00004$	&  $13.37\pm0.11$	&    $36.7\pm6.5$		\\ 
\ion{C}{IV}      	&  ...						&  $<12.9$	&    ...					\\ 
\ion{Si}{IV}      	&  ...						&  $<12.3$	&    ...					\\ 
\hline
\end{tabular}}
\end{center}
\end{table}

\begin{figure} 
\includegraphics[width=80mm]{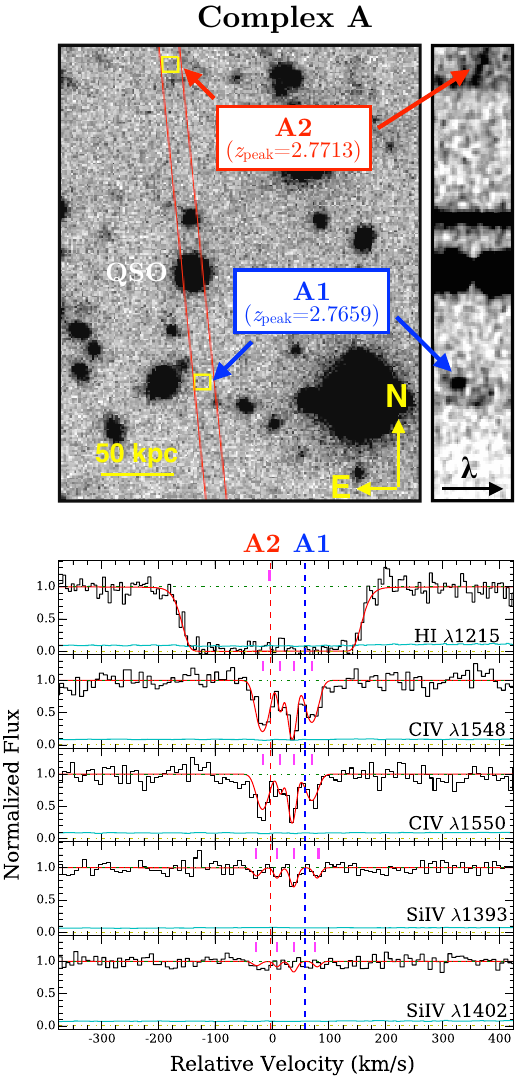}
\vspace{-0.5em}
\caption{
{\it Top-left}:  White-light image of the field around QSO DMS\,2139$-0405$. The slit orientation of the R08 long-slit observations is shown in parallel red lines. The positions of foreground LAEs A1 and A2 are indicated in yellow squares. The LAEs are situated on opposite sides of the QSO, with LAE A1 at $d=87$ kpc and LAE A2 at $d=161$ kpc. 
{\it Top-right}: A cutout of the two-dimensional long-slit spectrum, centered in spectral (horizontal) and spatial (vertical) directions at the Ly$\alpha$ absorption from absorption complex A. The red (blue) arrow points to the location of Ly$\alpha$ emission from LAE A1 (LAE A2).
{\it Bottom}: Continuum-normalized absorption profiles for various ionic transitions detected in absorption complex A. Zero velocity corresponds to the redshift of the Ly$\alpha$ absorption, $z=2.7691$. The absorption spectra and corresponding 1$\sigma$ error arrays are presented in black and cyan histograms. The dashed blue (red) line marks the adopted systemic redshift of LAE A1 (LAE A2)  (see \S 3). The best-fitting Voigt profile models are presented in red curves, with magenta tick marks indicating the centroid of individual components. For the saturated Ly$\alpha$ line, we estimate that the range of allowed $N\,\mathrm{(\ion{H}{I})}$ is $\log N\,\mathrm{(\ion{H}{I})}/ \cmjj=16.5-17.7$ at 95 percent confidence level, with a corresponding Doppler parameter range of $b=52-64$ \kms.}
\end{figure}

\begin{figure} 
\includegraphics[width=80mm]{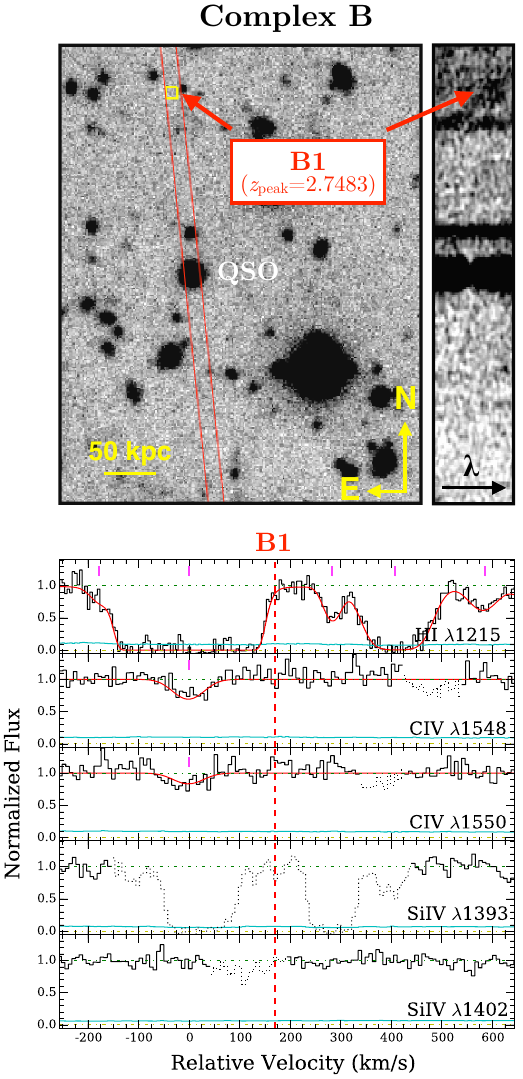}
\vspace{-0.5em}
\caption{
{\it Top-left}: White-light image of the field around QSO DMS\,2139$-0405$.  The yellow square indicates the position of foreground LAE B1 at $d=187$ kpc. The slit orientation of the R08 long-slit observations is shown in two parallel red lines.
{\it Top-right}: A cutout of the two-dimensional spectrum, centered in spectral (horizontal) and spatial (vertical) directions at the Ly$\alpha$ absorption from absorption complex B. The red arrow points to the location of Ly$\alpha$ emission from LAE B1. 
{\it Bottom}: Continuum-normalized absorption profiles of ionic transitions in absorption complex B. Zero velocity corresponds to the redshift of the strongest Ly$\alpha$ absorption component at $z=2.7440$. The absorption spectra and their corresponding 1$\sigma$ error array are shown in black and cyan histograms in each panel. Contaminating features are dotted out for clarity. The dashed red line marks the adopted systemic redshift of LAE 7. The best-fitting Voigt profile models are shown in red curves, whereas magenta tick marks indicate the location of individual absorption components. The strong Ly$\alpha$ line at 0 \kms is saturated. The estimated range of allowed $N\,\mathrm{(\ion{H}{I})}$ is $\log N\,\mathrm{(\ion{H}{I})}/ \cmjj=16.5-17.8$ at 95 percent confidence level , with a corresponding Doppler linewidth range of $b=48-61$ \kms.}
\end{figure}

\subsection[]{Gaseous environment of three $z=2.7-2.8$ LAEs}

As discussed in \S 3, three LAEs occur at $d<300$ kpc from the background QSO sightline (Figure 1). We now describe two Ly$\alpha$ absorption complexes which are coincident with these emitters in velocity space (see Figure 1). Each absorption system is found within a line-of-sight velocity difference of $|\Delta v| < 500$ \kms\ from the LAEs.

Complex A is found near the redshifts of LAE A1 and LAE A2 (see absorption panels in the bottom of Figure 3). Our MIKE echelle spectrum of  DMS 2139$-$0405 reveals strong \ion{H}{I} Ly$\alpha$ absorption with a total rest-frame equivalent width of $W_r(1215)=1.24\pm0.02$ \AA. Corresponding strong \ion{C}{IV} and modest \ion{Si}{IV} metal absorptions are also detected. The metal absorption profiles exhibit a multi-component structure that kinematically flank the systematic redshifts of the two LAEs. By fitting a single absorption component to the observed Ly$\alpha$ absorption, our Voigt profile fitting procedure returns a best-fit $\log N\mathrm{(\ion{H}{I})}/\cmjj=17.1\pm0.3$ for this system. The large uncertainty in $N\,\mathrm{(\ion{H}{I})}$ is understood as due to the saturated Ly$\alpha$ $\lambda$1215 profile and the fact that no higher-order \ion{H}{I} transition is available in the MIKE data. Based on $\chi^2$ values from the fitting analysis, we estimate that the range of allowed \ion{H}{I} column density is $\log N\mathrm{(\ion{H}{I})}/\cmjj=16.5-17.7$ at the 95 percent confidence level, with a corresponding Doppler parameter range of $b=52-64$ \kms. The range of allowed total $N\,\mathrm{(\ion{H}{I})}$ does not change significantly if we instead impose a multi-component H\,I profile that match the kinematic structure of the observed \ion{C}{IV} complex. 

The \ion{C}{IV} and \ion{Si}{IV} absorption profiles in Complex A can each be separated into a minimum of four discrete components which are spread over almost 100 \kms in line-of-sight velocity. We measure a total ionic column density of $\log N\mathrm{(\ion{C}{IV})}/\cmjj=14.21\pm 0.06$ for \ion{C}{IV} and $\log N\mathrm{(\ion{Si}{IV})}/\cmjj=12.84\pm0.09$ for \ion{Si}{IV}, integrated over all components. 

\begin{figure*} 
\includegraphics[width=180mm]{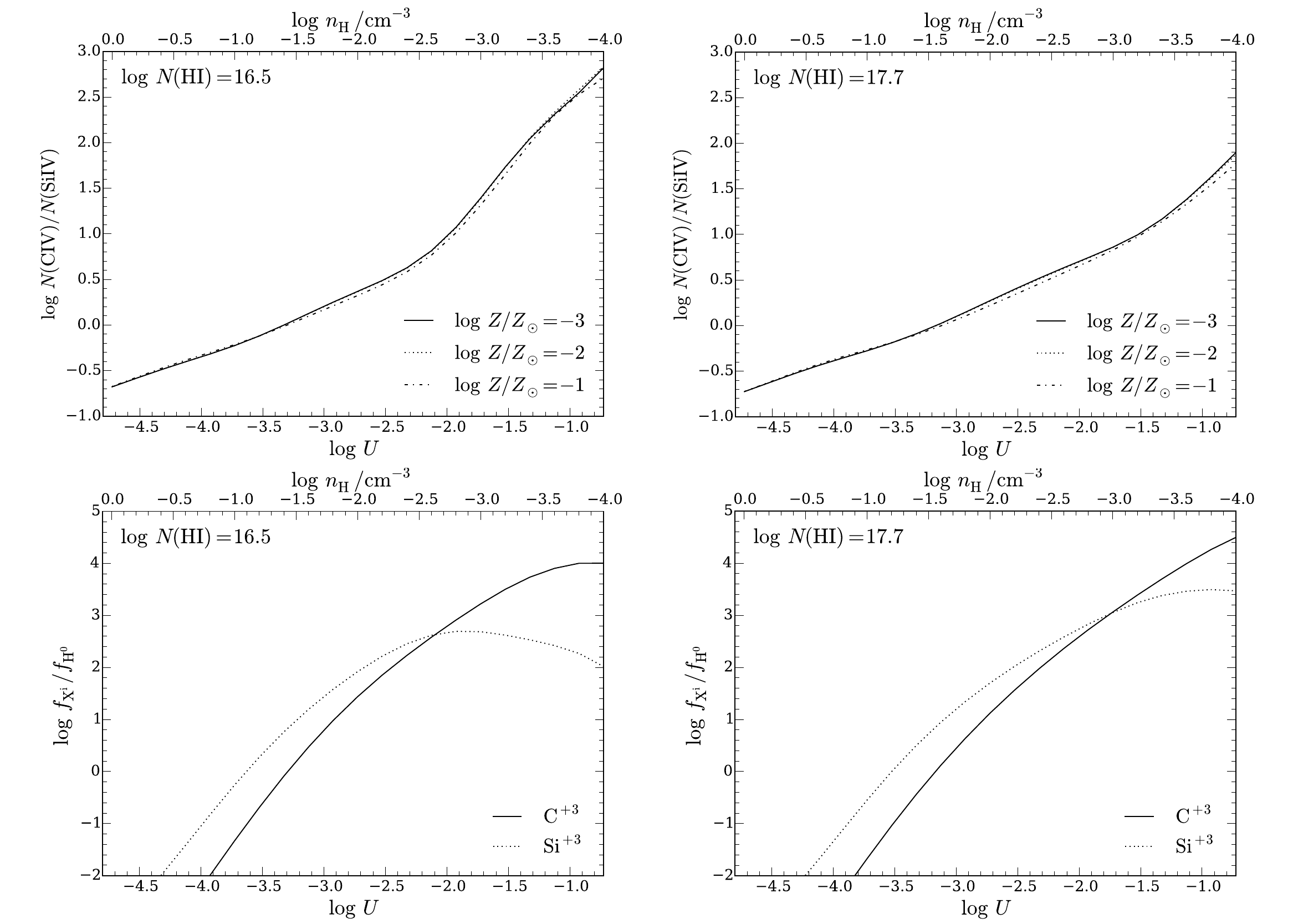}
\vspace{-1em}
\caption{Example of predictions from \textsc{cloudy} photoionization calculations. The top two panels show
the ionic column density ratio $ \log\,N\mathrm{(\ion{C}{IV})}/N\mathrm{(\ion{Si}{IV})}$ as a function of ionization parameter $U$. The calculations are performed for
for an optically thin gas ($\log N\mathrm{(\ion{H}{I})}/\cmjj=16.5$) on the left and optically thick gas ($\log N\mathrm{(\ion{H}{I})}/\cmjj=17.7$) on the right, matching the
allowed minimum and maximum $N\,\mathrm{(\ion{H}{I})}$ of the absorption complex A (see \S\ 4.1). In the bottom two panels, the ionization fractions of $\mathrm{C^{+3}}$ and $\mathrm{Si^{+3}}$ relative to $\mathrm{H^0}$ are plotted versus $U$ for both optically thin and thick cases.}
\end{figure*}

Complex B is found at the redshift of LAE B1. In contrast to the more kinematically compact absorption profile of Complex A, 
the Ly$\alpha$ profile of Complex B consists of multiple components spread over nearly 900 \kms\ in 
line-of-sight velocity (see the bottom panels of Figure 4). The integrated rest-frame equivalent width of the Ly$\alpha$ absorption is $W_r(1215)=1.99\pm0.03$ \AA. Our Voigt profile analysis shows that the total $N\,\mathrm{(\ion{H}{I})}$ in Complex B is dominated by a saturated Ly$\alpha$ component centered at $\Delta v = -170$ \kms\ blueward of the redshift of LAE B1. While the fitting procedure yields a best-fit $\log N\mathrm{(\ion{H}{I})}/\cmjj=17.3^{+0.3}_{-0.5}$ for this component, a full $\chi^2$ analysis indicates that the range of allowed column density is $\log N\mathrm{(\ion{H}{I})}/\cmjj=16.5-17.8$ at the 95 percent confidence level, with a corresponding Doppler linewidth of $b=48-61$ \kms. In addition to \ion{H}{I}, we detect modest \ion{C}{IV} absorption in this complex, at a location that matches the strongest Ly$\alpha$ component in velocity space. The best-fitting Voigt profile model for the \ion{C}{IV} absorption is given by a single-component with $ \log N\mathrm{(\ion{C}{IV})}/\cmjj=13.53\pm 0.07$. No \ion{Si}{IV} absorption is detected for the strongest Ly$\alpha$ component, for which we compute a 2$\sigma$ upper limit of $ \log N\,\mathrm{(\ion{Si}{IV})}<12.7$.

\subsection[]{The metallicity and ionization properties of the gas}

While our $N\mathrm{(\ion{H}{I})}$ measurements of the two gas absorption complexes are subject to large uncertainties because of the saturated Ly$\alpha$ absorption, the range of allowed $N\mathrm{(\ion{H}{I})}$ for each system is constrained to be log\,$N\mathrm{(\ion{H}{I})}/\cmjj \sim16-18$. 
Constraining \ion{H}{I} column density enables us to subsequently constrain the chemical enrichment level of the gas, based on the observed measurements or limits on the column density of ionic metals. Specifically, the mean elemental abundance of the gas, averaged across all components, is related to the column density ratio of the metal ions $\mathrm{C^{+3}}$ and $\mathrm{Si^{+3}}$ to $\mathrm{H^0}$, following an ionization fraction correction, 
$\log\,({\rm C}/{\rm H}) = \log\,N\mathrm{(\ion{C}{IV})}/N\mathrm{(\ion{H}{I})} -
\log\,\mathrm{(\mathit{f}_{C^{+3}}/\mathit{f}_{H^0})}$, and
$\log\,({\rm Si}/{\rm H}) = \log\,N\mathrm{(\ion{Si}{IV})}/N\mathrm{(\ion{H}{I})} -
\log\,\mathrm{(\mathit{f}_{Si^{+3}}/\mathit{f}_{H^0})}$,
where $f_\mathrm{H^{0}},\,f_\mathrm{C^{+3}}$, and $f_\mathrm{Si^{+3}}$ are the ionization fractions of H$^{0}$, C$^{+3}$, and Si$^{+3}$ ions, respectively.

For a photoionized gas, the ionization fractions of different species can be predicted given constraints on the ionization parameter of the gas, $U$,
which is defined as the number of incident ionizing photon per hydrogen atom. To estimate the necessary ionization fraction corrections, we perform a series of photoionization calculations using \textsc{Cloudy} v.13.03 (Ferland \etal\ 2013). We consider a plane parallel slab of gas which is irradiated with the updated Haardt \& Madau (2001) ionizing background radiation field (HM05 in \textsc{Cloudy}) at $z = 2.7$. We assume a solar chemical abundance
pattern for the gas, and that the gas is at ionization equilibrium. \textsc{Cloudy} then calculates the expected ionization fractions and column densities
for $\mathrm{H^0}$, $\mathrm{C^{+3}}$, and $\mathrm{Si^{+3}}$ ions over a wide range of gas densities and metallicities. Because the radiation field is fixed, changing the gas density leads to a change in the $U$ parameter. 

To estimate the metallicity of the gas, we perform two sets of  \textsc{Cloudy} calculations for each absorber: one for an optically thin gas ($\log N\mathrm{(\ion{H}{I})}/\cmjj<17.2$) and another for a gas that is optically thick ($\log N\mathrm{(\ion{H}{I})}/\cmjj>17.2$) to ionizing photons. The adopted $N\mathrm{(\ion{H}{I})}$ for these two sets of calculations are matched to the allowed minimum and maximum $N\,\mathrm{(\ion{H}{I})}$ value of each absorber (see \S\ 4.2). 
We present an example of \textsc{Cloudy} predictions in Figure 5, where in the top panels we plot the expected  $N\mathrm{(\ion{C}{IV})}/N\mathrm{(\ion{Si}{IV})}$ ratio plotted versus $U$ for a range of gas metallicities (from 0.001 to 0.1 solar metallicity). In this example, the calculation are performed for an optically thin gas with $\log N\mathrm{(\ion{H}{I})}/\cmjj= 16.5$ and optically thick gas with $\log N\mathrm{(\ion{H}{I})}/\cmjj=17.7$, bracketing the range of allowed $N\,\mathrm{(\ion{H}{I})}$ for Complex A. 

It is clear from the top two panels of Figure 5 that the predicted column density ratio of \ion{C}{IV} to \ion{Si}{IV} are insensitive to gas metallicity over the wide range of $U$ and metallicities probed, for optically thin and thick gases alike. Therefore, the ionization parameter of the gas can be constrained if the $N\mathrm{(\ion{C}{IV})}/N\mathrm{(\ion{Si}{IV})}$ ratio is known. Constraining $U$ then enables us to constrain the expected ionization fractions of $\mathrm{C^{+3}}$ and $\mathrm{Si^{+3}}$ relative to $\mathrm{H^0}$, as shown in the bottom panels of Figure 5.

For absorption complex A near LAEs A1 and A2, our Voigt profile fitting analysis yields a column density ratio of $\log N\mathrm{(\ion{C}{IV})}/N\mathrm{(\ion{Si}{IV})}=1.37\pm0.11$. Comparing the measured $N\mathrm{(\ion{C}{IV})}/N\mathrm{(\ion{Si}{IV})}$ to \textsc{Cloudy} predictions, we find that the estimated $U$ parameter of the gas ranges from as low as $\log U = -1.8 \pm0.1$ for the minimum allowed $N\,\mathrm{(\ion{H}{I})}$, $\log N\mathrm{(\ion{H}{I})}/\cmjj=16.5$, to possibly as high as $\log U = -1.2 \pm0.1$ for the maximum allowed \ion{H}{I} column density of $\log N\mathrm{(\ion{H}{I})}/\cmjj=17.7$. Adopting the range of allowed $U$ allows us to determine the relative ionization fractions of $\mathrm{C^{+3}}$ and $\mathrm{Si^{+3}}$ to $\mathrm{H^{0}}$. Adopting the solar chemical abundance pattern from Asplund \etal\ (2009), we estimate that the metallicity of the gas ranges from as low as $\mathrm{[M/H]} =-3.7\pm0.2$ for $\log N\mathrm{(\ion{H}{I})}/\cmjj=17.7$, to no higher than $\mathrm{[M/H]} =-1.8\pm0.2$ for $\log N\mathrm{(\ion{H}{I})}/\cmjj=16.5$.

For absorption complex B near LAE B1, the non-detection of \ion{Si}{IV} absorption implies a lower limit of $\log N\mathrm{(\ion{C}{IV})}/N\mathrm{(\ion{Si}{IV})}>0.83\pm0.07$. Based on our \textsc{Cloudy} calculations, the observed column density ratio requires $\log U > -2.1$, which in turn constrains the gas metallicity to 
$\mathrm{[M/H]} < -1.9$ for the allowed \ion{H}{I} column density range of $\log N\mathrm{(\ion{H}{I})}/\cmjj=16.5-17.8$.

It is instructive to compare the properties of these absorbers to similar \ion{H}{I} absorption systems  observed at $z=2-3$. For both absorption complexes A and B, the  the ionization parameter of the gas is constrained to $\log U\gtrsim-2$. This lower limit on $U$ is consistent with the observed ionization states of Lyman limit systems (LLSs; $\log N\mathrm{(\ion{H}{I})}/\cmjj>17.2$) at $z=2-3$, for which $\log U\gtrsim-3$ is common (e.g., Fumagalli \etal\ 2016a; Lehner \etal\ 2016). Furthermore, the allowed metallicities of the gas, which range from as low as $\sim0.001$ solar if the absorbers are optically thick, to no higher than $\sim0.02$ solar if the absorbers are in the optically thin regime, are similar to the the range of metallicities seen in $z=2-3$ LLSs (e.g., Fumagalli \etal\ 2016a; Lehner \etal\ 2016).\footnote{For our metallicity estimates, all the \ion{C}{IV} and \ion{Si}{IV} absorption are attributed to the total observed \ion{H}{I} absorption. It is possible that these ionic metals originate in only a small but metal-rich portion of the \ion{H}{I}-bearing gas complex. If such metal-rich pockets exist, the implication is that the large majority of the \ion{H}{I}-bearing gas would be even more metal-poor than inferred here, and the conclusion that the observed gas complex is predominantly metal-poor ($\lesssim0.01$ solar) would remain valid. In that regard, the metallicity constraints obtained in this section should be viewed as conservative upper limits to the mean chemical enrichment level in the bulk of the gas.}

\section[]{Discussion}

In the previous section, we characterize strong \ion{H}{I} absorption systems at $d<300$ kpc and $|\Delta v| < 500$ \kms\ from the three faint LAEs. It is also interesting to explore whether the gas is likely to be physically associated with these faint galaxies, and whether the galaxies themselves could be responsible for the origin of the gas. Because of the low completeness of the R08 spectroscopic observations of the field around DMS 2139$-$0405, it is challenging to link the absorption to a particular galaxy unambiguously, considering that it is possible other faint galaxies occur at smaller $d$ from the QSO than the nearest (in projection) LAE identified in our study (LAE A1 for Complex A  and LAE B1 for Complex B). For the sake of argument, however, we begin our discussion under the assumption that the absorption is associated with the CGM of the LAE that is closest to it in projected distance. We will put this assumption under a closer examination in the next subsection.

\subsection[]{Are the observed metals due to local outflows?}

We first investigate whether the gas can originate in a starburst-driven wind originating in the LAEs. Here we focus our discussion on Complex A, which is found closest to LAE A1 at $d=87$ kpc.  For a wind traveling with a velocity of $v_\mathrm{out}=200$ \kms, which is typical of $z\sim2.5$ LAEs (e.g., Trainor \etal\ 2015), a minimum of $\sim0.4$ Gyr is required for outflowing material to reach a distance of $\sim90$ kpc. This simple timing argument rests on the assumption that the outflow velocity can be maintained over cosmic time. In principle, if the galaxy had experienced a starburst episode at around $z\sim3.3$ or earlier, enough time would have elapsed for the outflowing material to reach the QSO sightline at the epoch of the observations. 

However, energetically it is a non trivial problem to propel gas to large distances from galaxies at high velocity (e.g., Gauthier \& Chen 2012). 
Moreover, while down-the-barrel spectroscopic observations of $z>2$ Lyman Break Galaxies (LBGs) and LAEs at have established the ubiquity of galactic-scale winds at high redshifts (e.g., Steidel \etal\ 2010; Hashimoto \etal\ 2013; Shibuya \etal\ 2014; Trainor \etal\ 2015), a primary caveat of the down-the-barrel method is that the galacto-centric distance of the gas is not well constrained. In the nearby universe, galactic-scale outflows have yet to be directly detected at $d\gtrsim10$ kpc from starburst galaxies. 

In spite of the lack of direct observation of outflowing gas far from galaxies, kinematic and chemical arguments have been put forward to link outflows with absorbing gas at $d\gtrsim50$ kpc from both low- and high-redshift galaxies (e.g., Borthakur \etal\ 2013; Crighton \etal\ 2015; Muzahid \etal\ 2015). In particular, outflowing gas is expected to exhibit a large velocity width and a high chemical enrichment level, due to the supernova origin of a super-galactic wind. Observations of blue-shifted ISM absorption lines in lensed LBGs at $z\sim2$ demonstrate that a high metallicity of $0.4-0.8$ solar is common in outflows (e.g., Pettini \etal\ 2002; Dessauges-Zavadsky \etal\ 2010). 

While no direct chemical abundance measurement is available for LAE A1, the lack of continuum counterpart of this LAE ($r>26.6$, $K_s>24.5$) implies that the host galaxy is intrinsically faint and low-mass ($L\lesssim0.1L^*$, $M_\mathrm{star} \lesssim10^9\,\mathrm{M_\odot}$). Based on the mass-metallicity relation at $z\sim 2$, the expected gas-phase metallicity in the ISM of a $M_*\sim10^8\,\mathrm{M_\odot}$ galaxy is $0.1-0.2$ solar (e.g., Steidel \etal\ 2014; Sanders \etal\ 2015). The expected ISM abundance is similar to the mean metallicity of $0.2\pm0.1$ solar estimated from nebular emission lines in stacked spectra of faint $z\sim2.5$ LAEs (Trainor \etal\ 2016). 

In contrast, the inferred metallicity of Complex A is $<0.02$ solar, which is at least an order of magnitude lower than both the observed metallicities in high-redshift outflows and the inferred chemical abundance in the ISM of faint LAEs.  The stark difference between the chemical enrichment level of the gas and those expected in outflowing material and ISM gas indicates that a starburst-driven wind is unlikely to be the origin of the observed gas.

Next, we examine whether the observed gas in Complex A is likely to be situated in the CGM of LAE A1. The total carbon mass enclosed within a projected distance $d$ in the CGM is $M_\mathrm{C} = \pi d^2 \langle N\mathrm{(\ion{C}{IV})}\rangle \, 12m_\mathrm{H} f_\mathrm{cov}/f_\mathrm{C^{+3}}$, where $\langle N\mathrm{(\ion{C}{IV})}\rangle$ is the mean \ion{C}{IV} column density, $f_\mathrm{cov}$ is the mean gas covering fraction and $f_\mathrm{C^{+3}}$ is the ionization fraction of triply ionized carbon. Assuming that that the radial profile of \ion{C}{IV} absorption within $d\sim90$ kpc has a mean column of $\langle N\mathrm{(\ion{C}{IV})}\rangle= 10^{14.2} \cmjj$ (which is the measured total \ion{C}{IV} column density in Complex A), and adopting $f_\mathrm{C^{+3}}=0.2$ based on our \textsc{cloudy} calculations, we estimate a total CGM carbon mass of $M_\mathrm{C} \sim 2\times10^6 \, \mathrm{M_\odot} \,(f_\mathrm{cov}) $. In a similar fashion, the observed $N\mathrm{(\ion{Si}{IV})}$  implies a total silicon mass of $M_\mathrm{Si} \sim 3\times10^5 \, \mathrm{M_\odot} \,(f_\mathrm{cov})$. For $f_\mathrm{cov}=0.5$, which is typical at similar $d$ from both low-redshift dwarf/sub-$L^*$ galaxies and high-redshift LBGs (e.g., Bordoloi \etal\ 2014; Liang \& Chen 2014; Rudie \etal\ 2019), the estimated total carbon and silicon masses in the CGM are $\sim 1\times10^6 \, \mathrm{M_\odot}$ and $\sim 1.5\times10^5 \, \mathrm{M_\odot}$, respectively. 
 
Can the galaxy produce the estimated amount of CGM metals? Assuming the galaxy has maintained a constant star-formation rate of $0.1\,\mathrm{M_\odot}\,\mathrm{yr^{-1}}$ (see \S\ 3) over time, and adopting a supernova carbon yield of 0.0088 $\mathrm{M_\odot}$ per solar mass of star formation (Peeples \etal\ 2014), we estimate that a total mass of $\sim 9\times10^4 \, \mathrm{M_\odot}$ in carbon is  produced over a $\sim0.1$ Gyr timescale. The recent metal production budget in the galaxy is far below the estimated total carbon mass to explain the observed metal enrichment of the surrounding IGM through local galactic feedback. This exercise suggests that the absorbing gas is unlikely to originate in any form of CGM around the individual LAEs.

\begin{figure} 
\includegraphics[width=84mm]{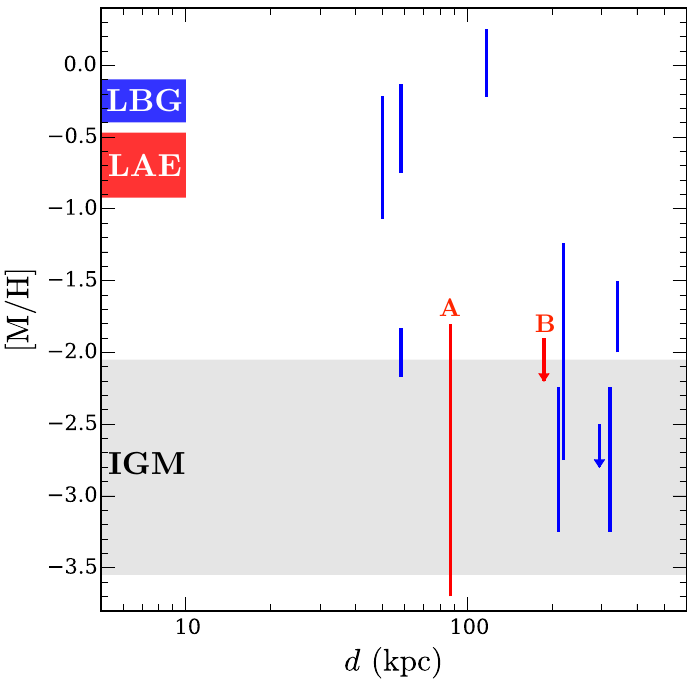}
\vspace{-1.6em}
\caption{Gas-phase metallicity measurements versus projected distance for $z>2$ galaxies. Constraints for gas metallicity in the vicinity of LAEs in our study are shown in vertical red bar and downward arrow, labeled by the absorption complex.  Note that while Complex A is shown here at $d=87$ kpc from LAE A1, another emitter LAE A2 is situated at $d=161$ kpc from the sightline. Meanwhile, the red rectangle at $d<10$ kpc marks the mean metallicity in the ISM of LAEs (Trainor \etal\ 2016), whereas the blue rectangle shows the typical ISM abundance level seen in LBGs (e.g., Pettini \etal\ 2002; Dessauges-Zavadsky \etal\ 2010; Strom \etal\ 2018). The range of metallicities seen in the $z\sim2.5$ IGM is shown in the shaded gray area (Simcoe \etal\ 2004). Other absorption line measurements at $d>10$ kpc from $z>2$ galaxies are shown in vertical blue bars (Simcoe \etal\ 2006; Crighton \etal\ 2013, 2015). The low metallicities seen at $d\sim100-200$\,kpc from faint LAEs in our study are consistent with the chemical enrichment level seen in the IGM at the same epoch.}
\end{figure}

\subsection[]{Evidence for accretion from a cosmic web filament?}

In Figure 6, we present gas-phase metallicity measurements versus $d$ for $z>2$ galaxies, including LAEs in our study. For comparison, we indicate the range of metallicities seen in the $z\sim2.5$ IGM in the shaded gray area (Simcoe \etal\ 2004).The inferred low metallicities of absorbers in our study (no higher than 0.02 solar and possibly lower than 0.001 solar metallicity) indicate that the gas originates in the more metal-poor IGM. Indeed, chemical enrichment level in the IGM at $z\sim2.5$ is characterized by a median metallicity of $\mathrm{[M/H]}=-2.8$ (or 0.002 solar) and a dispersion of 0.75 dex (e.g., Schaye \etal\ 2003; Simcoe \etal\ 2004), in excellent agreement with the metallicity constraints for absorbers in our study.  

Furthermore, the presence of multiple LAEs within a close proximity from each other along the slit implies that the emitters may be situated in a coherent structure. Using the field LAE luminosity function at $z\sim3$ from Drake \etal\ (2017), we expect to detect only $\sim0.3$ LAEs above the detection limit in the observations of R08, within a cosmological volume of $\sim30$ Mpc$^{3}$ defined by a slit width of 2 arcsec, a slit length of 453 arcsec, and a redshift range of $z=2.74-2.78$ ($\approx 3200$ \kms\ in velocity space, bracketing the redshifts of the Complex A and Complex B). Within this volume, the probability that three LAEs occur by random chance is very low, $\sim 3 \times 10^{-3}$. The probability that the grouping of three LAEs should occur within projected $\pm300$ kpc from each other is even lower, $\sim 1 \times 10^{-4}$, thereby suggesting that the LAEs are part of a spatially coherent structure. 
 
The most plausible explanation for the unique arrangement of Ly$\alpha$ emitters and absorbers in our study (Figure 1) is that the spectrograph slit in R08 was serendipitously aligned (at least partially) with a cosmic web filament at $z\sim 2.8$, thereby explaining the presence of multiple LAEs and Ly$\alpha$ absorbers along a narrow line extending more than 200 kpc in projection (see also Fumagalli \etal\ 2016b). Several features of the absorbing gas in Complex A near LAEs A1 and A2 provide further support for this scenario. 

First, the observed line-of-sight velocity spread of the multi-component \ion{C}{IV} absorption in Complex A is low, with a statistical dispersion of $\sigma=32$ \kms (see bottom panel of Figure 3). This velocity width is significantly smaller than those observed in down-the-barrel outflows from $z=2-3$ galaxies ($\gtrsim 100$ \kms, e.g., Trainor \etal\ 2015; Jones \etal\ 2018) and expected from dynamical motion in the CGM (the line-of-sight velocity dispersion for virialized motion is $\sigma_\mathrm{h}\approx60-130$ \kms\ in a $M_\mathrm{h}=10^{11-12}\,\mathrm{M_\odot}$ halo). However, the observed narrow \ion{C}{IV} width is consistent with the expected velocity spread for a quiescent, $\sim 100$ kpc scale gaseous filament that is simply moving with the Hubble flow at this redshift.\footnote{The Hubble parameter is $H=287\, {\rm km} \ {\rm s}^{-1}\ {\rm Mpc}^{-1}$ at $z=2.77$, for the adopted cosmology in this work.} Secondly, the absorption is flanked on opposite sides of the slit by the two LAEs (top panel of Figure 3), which are separated by a projected 248 kpc and are consistent within uncertainties with having zero systemic velocity offset from each other. Furthermore, the mean Doppler parameter for the \ion{C}{IV} absorption components in Complex A is $\langle b\rangle= 9.8\pm1.0$ \kms, which is very similar to the mode of the \ion{C}{IV} Doppler parameter distribution predicted by simulations of gas accretion in cosmic filaments (e.g., Rauch \etal\ 1997). Together, these kinematic and morphological features suggest that the absorbing gas originates in a large and dynamically cold structure, such as that of a gaseous filament. 

Recent hydrodynamical simulations of galaxy formation predict that cosmological accretion from the IGM onto high-redshift galaxies proceed along filaments that often manifest as low-metallicity LLSs (e.g., Fumagalli \etal\ 2011; van de Voort \etal\ 2012; Faucher-Gigu\`ere \etal\ 2015), in agreement with the observed properties of the absorption complexes in our study. In this physical picture, infalling streams of gas can proceed along such filamentary structure and fuel star-formation activity in the LAEs. This feeding scenario is further supported by the prominent blue peak that is observed in the Ly$\alpha$ emission line profile of LAE A1 (R08), an indication that the Ly$\alpha$-emitting gas is flowing onto the galaxy (e.g., Verhamme \etal\ 2006). In conclusion, the observed properties of the absorbers and the unique orientation of faint LAEs and the background sightline are in good agreement with the predictions of infalling IGM gas along cosmic filaments. 

\section[]{Summary}

Using high-resolution optical echelle spectrum of QSO DMS\,2139$-$0405 we studied the gaseous environment of three faint LAEs ($L_\mathrm{Ly\alpha}\approx 2 \times 10^{41}$ erg s$^{-1}$) at $z=2.7-2.8$. Strong \ion{H}{I} absorption systems are found at d $< 300$ kpc and $|\Delta v| < 500$ \kms\ from the LAEs, allowing us to investigate the connection between faint, low-mass galaxies ($L\lesssim0.1\,L^*$, $M_\mathrm{star}\lesssim10^9\,\mathrm{M_\odot}$) and the IGM at this redshift. Our findings are summarized below. 

\begin{enumerate}
\item We find that the absorbers have \ion{H}{I} column densities which are consistent with being partial or full LLSs, log\,$N\mathrm{(\ion{H}{I})}/\cmjj=16-18$. The observed metal abundances of the gas are low, no higher than 0.02 solar metallicity if the absorbers are optically thin and possibly as low or lower than $\sim0.001$ solar metallicity if the absorbers are optically thick LLSs. The inferred metallicities and ionization states of the gas are comparable to what have been observed in $z=2-3$ LLSs. 

\item Focusing on Complex A, which is found at $d=$ 87 and 161 kpc from LAEs A1 and A2, we examined the possibility that the gas originates in a locally enriched CGM around one of the galaxies. While a galactic wind with a constant velocity of 200 \kms\ may be formally able to cover a distance of $\sim 90$ kpc and reach the QSO sightline in $\sim0.4$ Gyr, direct evidence for the existence of outflows at large distances remains elusive (see \S\ 1). Furthermore, given the inferred low SFR of the LAE, the expected metal production budget is insufficient to explain the observed metals as having been produced and expelled by these galaxies within the last $\sim0.1$ Gyr.

\item The low metallicities of the gas are consistent with the gas originating in the general IGM. The spatial alignment of the LAEs along the slit, and the alignment of said LAEs and the absorbers in velocity space, suggest that we may be looking perpendicularly at a filamentary structure spanned by the galaxies over at least $\sim250$ kpc in projection. The range of $N\mathrm{(\ion{H}{I})}$ of the gas, its low metallicity and line-of-sight velocity dispersion, together with the observed mean \ion{C}{IV} Doppler parameter values and presence of faint galaxies in the vicinity of the absorbers, are all in good agreement with the predictions for infalling gas along cosmic filaments (Rauch, Haehnelt \& Steinmetz 1997; Fumagalli et al 2011, van de Voort et al 2012). The prominence of a blue Ly$\alpha$ emission peak in one of the LAEs (R08) is consistent with radiative transfer through infalling gas, making this a rare occasion where we can observe an accretion stream together with its final destination.
\end{enumerate}

We expect that future spectroscopic surveys of QSO fields in search of faint line emitters, taking advantage of wide-field integral field spectrographs and coupled with very deep imaging of these fields, will provide new insights into the gaseous environment of low-mass galaxies at high redshifts and offer critical information on the relative importance and roles of feeding and feedback during the cosmic high noon.

\section*{Acknowledgments}

FSZ acknowledges generous support from the Brinson Foundation and the Observatories of the Carnegie Institution for Science. AAS acknowledges support by the National Science Foundation under Grant AST1814719. We thank Andrew Bunker for kindly providing us with the reduced Keck LRIS images of DMS\,2139$-$0405. This work is based on data gathered with the 6.5 m Magellan Telescopes located at Las Campanas Observatory. Additional data were obtained at the W.M. Keck Observatory, which is operated as a scientific partnership among the California Institute of Technology, the University of California and the National Aeronautics and Space Administration.  The Observatory was made possible by the generous financial support of the W.M. Keck Foundation.

\bsp

\appendix
\section{Other  absorption systems at $z=2.7-2.8$.}

Two additional  Ly$\alpha$ absorption complexes are found at $z=2.7-2.8$ along the spectrum of  background QSO DMS\,2139$-$0405 (see Figure 1): Complex X at $z=2.7147$ and Complex Y at $z=2.7863$. No known LAE is found within $d<300$ kpc and $|\Delta v| < 500$ \kms\ from these Ly$\alpha$ absorption systems.

 A strong, saturated Ly$\alpha$ component dominates the total \ion{H}{I} column density of Complex X (Figure A1), with a best-fit $\log N\mathrm{(\ion{H}{I})}/\cmjj=17.9^{+0.2}_{-0.3}$. For this saturated system, we estimate that the range of allowed $N\,\mathrm{(\ion{H}{I})}$ is $\log N\mathrm{(\ion{H}{I})}/\cmjj=15.7-18.2$ at 95 percent confidence level. In addition to Ly$\alpha$, \ion{C}{IV} absorption is observed in Complex X as well, which corresponds with the strongest Ly$\alpha$ absorption component in velocity space, with a measured total column density of $ \log N_\mathrm{tot}\,\mathrm{(\ion{C}{IV})}=13.84\pm 0.11$. Unfortunately, it is not possible to constrain the \ion{Si}{IV} column density, due to contamination from Ly$\alpha$ forest lines at higher redshifts. 

Only Ly$\alpha$ absorption is detected in Complex Y (Figure A2). The range of allowed $N\,\mathrm{(\ion{H}{I})}$ for this saturated system is $\log N\mathrm{(\ion{H}{I})}/\cmjj=16.7-17.8$ at 95\% confidence level. For the absence of ionic metal absorption, we calculate upper limits on $ \log N\,\mathrm{(\ion{C}{IV})}/\cmjj<13.2$ and $ \log N\,\mathrm{(\ion{Si}{IV})}/\cmjj<12.6$ for \ion{C}{IV} and \ion{Si}{IV}, respectively.

\begin{figure} 
\includegraphics[width=80mm]{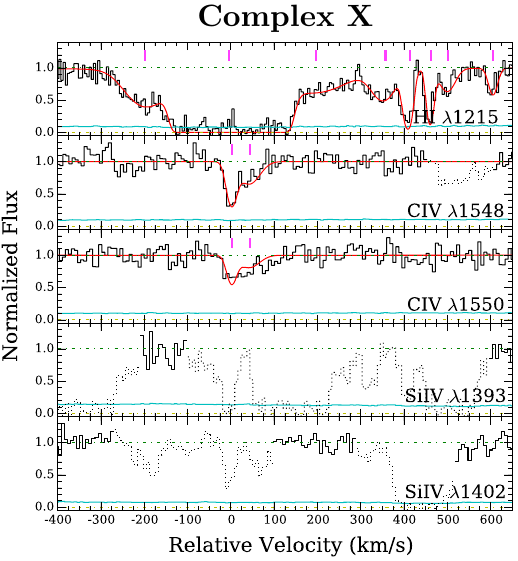}
\vspace{-1em}
\caption{Continuum-normallized absorption profiles of Complex X at $z=2.7147$. The absorption spectra and
1$\sigma$ error array are shown in black and cyan histograms, whereas the best-fitting Voigt profile models are shown in red curves.
The magenta tick marks indicate the location of individual components. Contaminating features are dotted out for clarity.}
\end{figure}

\begin{figure} 
\includegraphics[width=80mm]{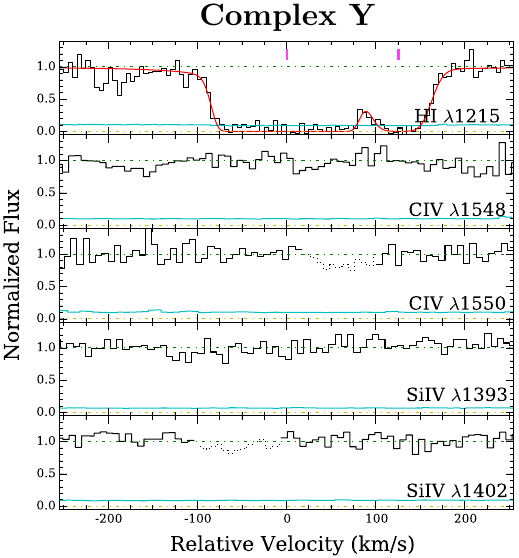}
\vspace{-1em}
\caption{Continuum-normalized absorption profiles of Complex Y at $z=2.7863$. The absorption spectra and
1-$\sigma$ error array are shown in black and cyan histograms, whereas the best-fitting Voigt profile models are shown in red curves.
The magenta tick marks indicate the location of individual components. Contaminating features are dotted out for clarity.}
\end{figure}

\begin{table}
\begin{center}
\caption{Voigt profile fitting results for additional absorption complexes at $z=2.7-2.8$} 
\vspace{-0.5em}
\label{tab:Imaging}
\resizebox{3.4in}{!}{
\begin{tabular}{cccc}\hline
Species		& $z_c$ 					& log\,$N_c$/\cmjj		&$b_c$\\	
 		&								&    		   	& (km\,s$^{-1}$)            \\\hline \hline

\multicolumn{4}{c}{Complex X at $z=2.7147$}\\\hline
\ion{H}{I}      	&   $2.7122\pm0.0001$	&  $13.82\pm0.12$	&  $57.0\pm10.7$	\\ 
\ion{C}{IV}      	&  ...						&  $<13.0$	&    ...					\\ 
\ion{Si}{IV}      	&  ...						&  $<12.7$	&    ...					\\ 
\ion{H}{I}      	&   $2.71466\pm0.00003$	&  $17.9^{+0.2}_{-0.3}$	&  $47^{+2}_{-1}$	\\ 
\ion{C}{IV}    	&   $2.71469\pm0.00003$	&  $13.58\pm0.14$	&  $13.7\pm3.9$	\\
\ion{C}{IV}     	&   $2.71519\pm0.00015$	&  $13.50\pm0.18$	&  $31.8\pm13.5$	\\ 
\ion{H}{I}      	&   $2.71710\pm0.00037$	&  $13.73\pm0.25$	&  $94.3\pm44.3$	\\ 
\ion{C}{IV}      	&  ...						&  $<13.1$	&    ...					\\ 
\ion{H}{I}      	&   $2.71904\pm0.00005$	&  $13.50\pm0.09$	&  $35.0\pm8.0$	\\ 
\ion{C}{IV}      	&  ...						&  $<12.9$	&    ...					\\ 
\ion{H}{I}      	&   $2.71974\pm0.00001$	&  $13.71\pm0.13$	&  $11.2\pm2.0$	\\ 
\ion{C}{IV}      	&  ...						&  $<12.7$	&    ...					\\ 
\ion{H}{I}      	&   $2.72037\pm0.00002$	&  $13.49\pm0.12$	&  $8.3\pm2.0$	\\ 
\ion{C}{IV}      	&  ...						&  $<12.9$	&    ...					\\ 
\ion{H}{I}      	&   $2.72086\pm0.00006$	&  $13.15\pm0.11$	&  $23.0\pm7.1$	\\ 
\ion{C}{IV}      	&  ...						&  $<13.1$	&    ...					\\ 
\ion{H}{I}      	&   $2.72217\pm0.00003$	&  $12.99\pm0.10$	&  $12.3\pm3.8$	\\ 
\ion{C}{IV}      	&  ...						&  $<13.0$	&    ...					\\ 
\ion{Si}{IV}      	&  ...						&  $<12.4$	&    ...					\\  \hline

\multicolumn{4}{c}{Complex Y at $z=2.7863$}\\\hline

\ion{H}{I}    	&   $2.78630\pm0.00002$	&  $17.6^{+0.1}_{-0.6}$&    $27^{+3}_{-1}	$	\\
\ion{C}{IV}      	&  ...						&  $<12.9$	&    ...					\\ 
\ion{Si}{IV}      	&  ...						&  $<12.3$	&    ...					\\ 
\ion{H}{I}     	&   $2.78790\pm0.00002$	&  $14.34\pm0.15$	&  $24.0\pm3.2$	\\ 
\ion{C}{IV}      	&  ...						&  $<12.8$	&    ...					\\ 
\ion{Si}{IV}      	&  ...						&  $<12.2$	&    ...					\\

\hline

\end{tabular}}
\end{center}
\end{table}

\label{lastpage}

\end{document}